\documentclass[sigconf, nonacm]{acmart}
\usepackage{colortbl}
\usepackage{subcaption}
\usepackage{tabularx}
\usepackage{csquotes}
\usepackage{booktabs}
\usepackage{array}
\usepackage{graphicx}
\newcommand{\insertquote}[2]{
    \begin{displayquote}
    \parbox{\linewidth}{\textit{#1} [#2]}
    \end{displayquote}
}

\newcommand{\participant}[1]{%
    \ifnum #1=1\text{L1}\else\ifnum #1=2\text{L2}\else\ifnum #1=3\text{E1}\else\ifnum #1=4\text{E2}\else\ifnum #1=5\text{L3}\else\ifnum #1=6\text{E3}\else\ifnum #1=7\text{L4}\else\ifnum #1=8\text{L5}\else\ifnum #1=9\text{L6}\else\ifnum #1=10\text{E4}\else\ifnum #1=11\text{L7}\else\ifnum #1=12\text{E5}\else\ifnum #1=13\text{E6}\else\ifnum #1=14\text{E7}\else\ifnum #1=15\text{E8}\else\ifnum #1=16\text{E9}\else\ifnum #1=17\text{E10}\fi\fi\fi\fi\fi\fi\fi\fi\fi\fi\fi\fi\fi\fi\fi\fi\fi
}

\newcommand{\review}[1]{\textcolor{black}{#1}}

\AtBeginDocument{%
  }

\begin{document}
% \SetWatermarkScale{0.5}
\title{Incorporating Sustainability in Electronics Design: Obstacles and Opportunities}

\author{Zachary Englhardt}
\email{zacharye@cs.washington.edu}
\orcid{0000-0002-6646-6466}
\affiliation{%
  \institution{University of Washington}
  \country{Seattle, Washington, USA}
}

\author{Felix H{\"a}hnlein}
\email{fhahnlei@cs.washington.edu}
\orcid{0000-0002-3484-4004}
\affiliation{%
  \institution{University of Washington}
  \country{Seattle, Washington, USA}
}

\author{Yuxuan Mei}
\email{ym2552@cs.washington.edu}
\orcid{0000-0003-1024-9267}
\affiliation{%
  \institution{University of Washington}
  \country{Seattle, Washington, USA}
}

\author{Tong Lin}
\email{tlin01@cs.washington.edu}
\orcid{0009-0007-2267-6377}
\affiliation{%
  \institution{University of Washington}
  \country{Seattle, Washington, USA}
}

\author{Connor Masahiro Sun}
\email{suncon@oregonstate.edu}
\orcid{0009-0001-0437-6174}
\affiliation{%
  \institution{Oregon State University}
  \country{Corvallis, Oregon, USA}
}

\author{Zhihan Zhang}
\email{zzhihan@cs.washington.edu}
\orcid{0000-0001-7394-5409}
\affiliation{%
  \institution{University of Washington}
  \country{Seattle, Washington, USA}}

\author{Shwetak Patel}
\email{shwetak@cs.washington.edu}
\orcid{0000-0002-6300-4389}
\affiliation{%
  \institution{University of Washington}
  \country{Seattle, Washington, USA}}

\author{Adriana Schulz}
\email{adriana@cs.washington.edu}
\orcid{0000-0002-2464-0876}
\affiliation{%
  \institution{University of Washington}
  \country{Seattle, Washington, USA}}

\author{Vikram Iyer}
\email{vsiyer@uw.edu}
\orcid{0000-0002-3025-7953}
\affiliation{%
  \institution{University of Washington}
  \country{Seattle, Washington, USA}}

\renewcommand{\shortauthors}{Englhardt et al.}

\newcommand{\todo}[1]{{\color{red}{TODO: #1}}}
\newcommand{\meta}[1]{}

\newcommand{\takeaway}{\noindent\textbf{Key Takeaways: }}

\begin{abstract}
%VI: I tried to explicitly mention that data gathering is a key challenge and be more explicit that the HCI community (or HCI technologies for -1 word) can address them 

Life cycle assessment (LCA) is a methodology for holistically measuring the environmental impact of a product from initial manufacturing to end-of-life disposal.
However, the extent to which LCA informs the design of computing devices remains unclear.
To understand how this information is collected and applied, we interviewed 17 industry professionals with experience in LCA or electronics design, systematically coded the interviews, and investigated common themes. 
These themes highlight the challenge of LCA data collection and reveal distributed decision-making processes where responsibility for sustainable design choices—and their associated costs—is often ambiguous.
Our analysis identifies opportunities for HCI technologies to support LCA computation and its integration into the design process to facilitate sustainability-oriented decision-making.
While this work provides a nuanced discussion about sustainable design in the information and communication technologies (ICT) hardware industry, we hope our insights will also be valuable to other sectors.

\end{abstract}





\maketitle

\section{Introduction}
%Adriana's version
%The environmental impact of electronics production is enormous and growing rapidly. Electronics are integral to modern life, but their production contributes significantly to global emissions and waste. Recent studies found that in 2022, the world generated 62 billion kg of e-waste and only 22.3\% were properly collected and recycled~\cite{balde2024global}. Furthermore, electronics manufacturing is responsible for nearly 4\% of global greenhouse gas emissions~\cite{freitag_real_2021}, and this figure is expected to rise~\cite{andrae2015global}. As consumer demand for technology increases, the environmental toll of electronics production continues to be a pressing and urgent problem. %\todo{fix data reports and add citations.}

%VI revised
The proliferation of computing into almost every aspect of modern life has led to substantial growth in its environmental impact (EI). Estimates show that the impact of computing is on par with the airline industry, accounting for up to 2.1-3.9\% of global climate warming emissions today~\cite{freitag_real_2021} and this impact is projected to double within the next decade~\cite{andrae2015global}. While transitioning to clean energy sources is critical, this alone will not address the 50-80\% of computing's carbon emissions that come from their resource intensive manufacturing~\cite{gupta_act_2022, gupta2021chasing, lyu2023myths, apple-device-lca}. It will also not address the annual 62 million metric tons of electronic waste (e-waste) they produce at end of life~\cite{e-waste2024}. Considering many computing devices have lifetimes of a few years, it is imperative that we begin designing the next generation of sustainable devices \textit{now} to achieve environmental goals such as net zero emissions set by many organizations for 2030 and beyond.

% Adriana's version
%Despite growing interest from various stakeholders to reduce the environmental impact of electronics, significant challenges remain. Life-Cycle Assessments (LCA) is a key tool in evaluating environmental impact, but they are costly and computationally complex. Comparing LCAs across different suppliers and manufacturers can be difficult due to inconsistent data and varying methodologies. Additionally, integrating LCA results into the design process presents further obstacles, as designers often lack tools that can effectively translate these assessments into actionable insights.

%VI revised
\review{Addressing sustainability in product development presents two interlinked challenges: accurately assessing environmental impact throughout a product's life cycle and effectively incorporating these insights into the design process itself. This impact assessment is }typically done through a comprehensive LCA. This process is both expensive and time-consuming. A sustainability expert traces each of the subcomponents in a device back to its raw materials and manufacturing energy inputs. The expert also assesses the device's energy consumption during use and the impacts of its end-of-life disposal. While complete LCAs provide important, detailed insights, in electronics they are often used as reporting tools for retrospective analysis rather than informing actionable decisions during the design process. \review{This area is ripe for exploration as a new avenue for the HCI community to advance sustainability by building systems that bridge the gap between complex LCA data and design decision-making.}

\review{In this paper, we seek to understand the interplay between current LCA and product development practices, arming researchers with the necessary background to drive meaningful change. }
We conducted a series of semi-structured interviews with professionals who have hands-on experience in producing LCAs or in designing electronics, giving the CHI research community a window into how LCAs are currently conducted \review{and are (or are not) used to inform design decisions,} where the key challenges lie, and what opportunities exist for future work. In total, we interviewed 17 experts, including LCA engineers, chief scientists, CEOs of LCA firms, semiconductor manufacturers, consortium managers, firmware and electrical engineers, and program managers.
% Researchers must first understand the diverse needs of the different stakeholders to 

Our analysis revealed several key themes, which we grouped into four categories: (1) the current LCA practices in the ICT sector; (2) the intersection of sustainability and product development; (3) navigating the product design and LCA ecosystem (4) incentive structures between stakeholders. We found that the design process is distributed across various domains of expertise, with environmental impact reporting often treated as a separate, external process that struggles to influence the core of product development. Moreover, it remains unclear who within the design and manufacturing pipeline should be responsible for making sustainable design decisions.

\review{We envision a future ecosystem of computational tools that addresses these challenges by accelerating the LCA process and presenting information about the environmental impact of computing devices in a format that is \textit{actionable} for designers and other decision-makers}. We observe that the CHI community is particularly well-positioned to study and develop innovative tools that support sustainable design practices, since both the LCA process and product development involve coordinating across a diverse range of stakeholders, including engineers, designers, sustainability analysts, fabrication experts, and a global supply chain. Developing new interaction paradigms for decision-makers and designers raises critical framing questions, such as identifying the users and their ecological context~\cite{bardzell2011towards}, examining power dynamics within the product development decision-making process, and bridging information gaps to produce and utilize insights gained from LCAs.

% This would be the first step toward creating computational design tools that can optimize and support the complex trade-offs in sustainable design. For example, is it better to choose an improved processor with lower energy consumption, or an older generation with lower embodied carbon from manufacturing? Extending this idea further to make carbon labels as ubiquitous as food nutrition labels could empower consumers to make environmentally conscious choices and even motivate policy decisions.

In this work, we contribute (1) \review{a series of interviews with LCA practitioners and engineers to provide insight into how LCAs are conducted in the ICT sector and the ways product designers interact with these reports,} and (2) opportunities for future research into developing new methods and computational tools for both LCA practitioners and engineers. These opportunities could significantly enhance the integration of sustainability into the electronics design process, providing practical pathways for making informed, environmentally conscious decisions. We hope this paper will spark new research efforts and support the development of impactful systems that contribute to reducing the environmental footprint of electronics.

\section{Related Work and Background}
\label{sec:related_work}
\review{This paper analyzes the challenges and opportunities associated with incorporating LCA methodologies into the electronics design process, taking inspiration from prior HCI works focusing on understanding the workflows of professional groups~\cite{khurana_adaptivesports_2021, sterman_creative_2022, LAKA2022100652} and electronics manufacturing~\cite{khurana2020beyond, zhu2022scaling}.} We review relevant literature on sustainable HCI and electronic design tools for electronics. We then give a brief primer on LCA, an overview of the landscape of tools to support LCA in electronics,  and situate our work in this context.

\subsection{Sustainable HCI}
Sustainability has been a growing focus in HCI for over a decade\review{~\cite{mencarini_imagining_2024, yan_future_2023, hansson_decade_2021}, with significant attention given to developing persuasive technologies to encourage sustainable behaviors and choices. These technologies generally seek to act as decision support tools or lifestyle intervention systems. A substantial body of work has explored creating tools to support sustainable choices by providing actionable environmental impact information in specific domains~\cite{ericsson_role_2020}. Notable examples include climate conscious travel planners that highlight lower-emission routes~\cite{google_check_2024}, sustainable shopping assistants~\cite{busheska_enroute_2024}, and carbon footprint calculators for food choices~\cite{my_emissions_food_2024}. While these tools cannot replace comprehensive LCA studies, they seek to offer accessible and interpretable estimates that enable users to make more environmentally-conscious decisions. }

\review{Much of the recent hardware-centered approaches in the HCI community have focused on incorporating biodegradable materials and recyclable components. Researchers have explored constructing interactive interfaces out of biodegradable materials~\cite{zhang_biodegradable_2024, koelle_prototyping_2022}. Others have proposed bio-hybrid approaches such as mycelium-based bio-hybrid devices~\cite{vasquez_myco-accessories_2019, lu_living} that incorporate living organisms rather than relying solely on traditional computing hardware. This represents a fundamental shift in hardware design, moving from traditional persistent materials to ones that can safely return to the environment at the end of their useful life. Parallel to these biodegradable and bio-hybrid solutions, significant work has focused on developing electronic systems that are more easily recycled~\cite{zhang_recyclable_2024, cheng_recy-ctronics_2024, yan_solderlesspcb_2024}. These advances are complemented by the development of novel design tools that incorporate sustainability considerations from early stages of product development. Recent works in this space include EcoEDA~\cite{lu_ecoeda_2023} which assists designers in sourcing used parts for new designs, as well as DeltaLCA~\cite{zhang_deltalca_2024} and EcoSketch~\cite{chatty_ecosketch_2024}, which seek to enable environmental impact assessment early on in the design process. }

\review{The sustainable HCI (SHCI) community has maintained an active dialogue of self-reflection and critique of the progress and impact of SHCI research~\cite{bremer_how_2023, bremer_have_2022, soden_what_2021}. Several researchers have raised important concerns about the effectiveness and limitations of current approaches, particularly regarding the rebound effects of implementing new technologies and the lack of practical, actionable steps for implementing sustainable solutions~\cite{silberman_next_2014, brynjarsdottir_sustainably_2012}. Remy et al.~\cite{remy_evaluation_2018, remy_limits_2017} have specifically addressed the need for better evaluation methods in SHCI research, highlighting the importance of developing more rigorous approaches to assess the effectiveness of sustainable computing initiatives. }

\review{Translating these research innovations to widespread industry adoption remains challenging~\cite{liu_out_2018, knowles_this_2018}. This gap between academic research and industry practice is particularly evident in sustainable computing, where theoretical solutions and research prototypes often struggle to find paths to widespread adoption. The limitations identified above in academic critique\textemdash particularly around impact measurement and actionable steps\textemdash closely mirror the challenges we observed in our interviews with industry practitioners. By shedding light on these practical obstacles, our work seeks to support the SHCI research community in effectively applying their expertise to real-world sustainability challenges in the technology sector. }

\subsection{\review{HCI for Electronics Design}}
Modern computing devices consist of of numerous interconnected components, ranging from simple passive components such as capacitors and resistors to complex silicon-based integrated circuits, or ICs, which can include more advanced functionality like digital logic, memory, and power regulation. These components are assembled on a printed circuit board (PCB), which consists of an insulating backing and conductive metal interconnects to which individual components are attached. Designers rely extensively on electronic design automation (EDA) tools like Altium Designer~\cite{altium-altium} and KiCad EDA~\cite{kicad_kicad_2024}, which streamline core aspects of the design process, including schematic creation, simulation, PCB layout, and the generation of design files and bill of materials (BOM) necessary for manufacturing. Interaction and design research in this space is highly active. Projects such as Polymorphic Blocks~\cite{lin-polymorphic} aim to exploit programming language principles such as polymorphism and encapsulation to enable \review{re-use of circuit modules across designs. Others such as Strasnick et al~\cite{strasnick-scanalog} }have focused on the iterative nature of electronics design, developing tools to support interactive, context-aware circuit debugging. 

Building off these efforts to incorporate EDA tools more holistically within the design process, \review{researchers have begun to explore methods to bridge the gap between prototyping and manufacturing. Through a qualitative study of the experiences of low-volume hardware designers, Khurana et al.~\cite{khurana2020beyond} revealed key challenges in transitioning from prototype to mass production, highlighting how early design decisions in prototyping can significantly impact manufacturability, and emphasizing the need for tools that consider manufacturing constraints from the outset. These findings have led to tools such as MakeDevice~\cite{kobi_make}, which assists in generating production-ready designs from JacDac~\cite{ball_jacdac} system prototypes. Our work is inspired by this need-finding approach, and we similarly highlight the potential for new tools to empower electronics designers to consider the downstream impacts of design decisions.  }

\subsection{\review{Life Cycle Assessment}}
\label{sec:lca-desc}
% - what is LCA
LCA is a process for systematically estimating the total EI of a product. LCA can be roughly broken into two steps: life cycle inventory (LCI) and environmental impact assessment (EIA). Creating the LCI involves decomposing a device into subcomponents to identify all of the inputs (e.g., raw materials, natural resources, energy), outputs (e.g., intermediate and final manufactured products), and the mapping of these quantities to potential EIs (e.g., carbon emissions, environmental toxicity) of a product throughout its life cycle~\cite{suh_methods_2005}. Then in the EIA stage, analysts construct a model of the production process flow, usage and disposal, and use the data obtained from LCI to compute the environmental impact estimates. 

\review{Once the LCA has been completed, a growing number of companies release consumer-facing product environmental impact reports, which tabulate the results of the LCA to report metrics such as carbon footprint numbers. In the ICT industry, this process is typically completed \textit{after} beginning production. To further complicate matters, the information reported in an LCA does not translate into clear design recommendations. Bhander et al.~\cite{bhander_implementing} describe this as the environmentally-conscious design process paradox, where conducting a LCA is contingent upon first completing the design, at which point it is often infeasible to make design changes based on the results of the analysis. }

To address this challenge, the construction and architecture sectors have begun placing a consistent focus on early-stage design optimization towards more sustainable building designs~\cite{stazi_life_2012, mendez_echenagucia_early_2015, hegazy_multi-objective_2021, mendez_echenagucia_tradeoffs_2023}. \review{This has led to widespread integration of LCA estimation and simulation functionality directly into building information management (BIM) software, or computational tools that enable designers to explore the complex trade-offs between different building envelope designs~\cite{HOLLBERG_BIM}. Having access to this information during the early design stage is invaluable due to the significant cost of building construction.}
\review{The computing and electronics industry faces similar challenges, however, there is currently a lack of support for early-stage EI estimation of electronics designs. Recent work has begun to explore this topic with researchers in computer architecture and systems developing guidelines and carbon footprint estimation tools~\cite{gupta_act_2022} with a focus on modeling the EI of data centers and cloud computing systems~\cite{acun_carbon_2023, patel_hotcarbon_2023, wang_peeling_2023, elgamal_carbon-efficient_2023}. Similarly, researchers have developed a handful of tools to evaluate aspects of PCB designs for sustainability~\cite{chatty_ecosketch_2024,zhang_deltalca_2024}. }%perhaps something about how these aren't widely adopted/used yet? 

Throughout the rest of this paper, we seek to support this nascent body of work through a comprehensive analysis of domain specific challenges in performing LCA for electronics and integrating these insights into the product design process. We identify these challenges in Section~\ref{sec:findings}, and in Section~\ref{sec:opportunities} we present a set of opportunities for the CHI community to address them by developing tools to support the integration of LCA methodologies into the design of electronic and computing systems.

\section{Methods} 
We compiled a list of 15 questions designed to uncover the challenges of producing LCAs of consumer electronic devices and how EI data is currently employed by engineers who design these products. 
We began by conducting two pilot interviews with hardware engineers who design PCBs for consumer electronics devices. These pilot interviews suggested that a rigid question set was insufficient; participants' varying roles and company structures necessitated a more adaptable approach. We instead adopted a semi-structured interview process using our initial compiled questions as a framework to guide the discussion similar to prior works~\cite{khurana2020beyond, khurana_adaptivesports_2021, zhu2022scaling, sterman_creative_2022, LAKA2022100652} that identify challenges and opportunities for professional groups.

\begin{table*}[t]
\centering
\resizebox{\textwidth}{!}{%
\begin{tabular}{@{}p{1.5cm}p{4.5cm}p{4cm}p{1.5cm}p{2.5cm}p{1cm}p{1cm}@{}}
\toprule
\textbf{Identifier} & \textbf{Role} & \textbf{Industry} & \textbf{Category} & \textbf{Years Experience} & \textbf{Region} & \textbf{Gender} \\ \midrule
\rowcolor{gray!15} L1  & Chief Scientist & Life Cycle Assessment & LCA        & 35 & EU & M \\ 
L2  & Chief Expert Environmental Protection Technology & ICT & LCA        & 30 & EU & M \\ 
\rowcolor{gray!15} L3  & LCA Engineer & Cloud/Data Center & LCA        & 10 & NA & M \\ 
L4  & LCA Expert & Electronics & LCA        & 10 & EU & M \\ 
\rowcolor{gray!15} L5  & Consortium Manager & Semiconductor Manufacturing & LCA        & 3  & NA & M \\ 
L6  & CEO & LCA Software & LCA        & 7  & NA & M \\ 
\rowcolor{gray!15} L7 & Researcher & Cloud Computing & LCA        & 2  & NA & M \\ 
E1  & Sr. Director Systems Engineering & Semiconductor Manufacturing & Engineer   & 42 & NA & M \\ 
\rowcolor{gray!15} E2  & Dram Design Engineer & Semiconductor Devices & Engineer   & 3  & NA & M \\ 
E3  & Systems Engineer & Mobile ICs & Engineer   & 2  & NA & M \\ 
\rowcolor{gray!15} E4 & Electrical Engineer & Consumer Electronics & Engineer   & 10 & NA & M \\ 
E5 & Firmware Engineer & Display Electronics & Engineer   & 3  & NA & M \\ 
\rowcolor{gray!15} E6 & Firmware/Hardware Engineer & Consumer Electronics & Engineer   & 10 & NA & M \\ 
E7 & Engineer / Researcher & Augmented Reality & Engineer   & 8  & NA & M \\ 
\rowcolor{gray!15} E8 & Senior Researcher & Computing Research & Engineer   & 5  & NA & W \\ 
E9 & Program Manager & Computer Technology - Software/Firmware/Manufacturing & Engineer   & 5  & NA & M \\ 
\rowcolor{gray!15} E10 & Application Engineer & Semiconductor Devices & Engineer   & 2  & NA & M \\ 
\bottomrule
\end{tabular}%
}
\vspace{0.3cm}
\caption{\textbf{Participant information.} 
Table displaying the breakdown of the participant's self-reported job title, industry, years of experience, region, and gender. Based on the participant's reported job responsibilities, we further categorize them as either LCA professionals or engineers. Each participant is linked with an identifier code in the left column, with codes starting with L denoting LCA professionals and codes starting with E denoting engineering and product development professionals.}
\end{table*}
\subsection{Study Procedure}
We conducted 17 interviews to understand the current practices when conducting LCAs of computing devices and the current role of sustainability considerations in the product development process. The interviews were conducted over Zoom video teleconferencing. Interviews were scheduled for one hour. In each semi-structured interview, we began by asking the participant to describe their current role and responsibilities in detail. Next, we asked them to describe a timeline and their contributions in conducting an LCA or working on a product development cycle. We then transitioned to discussing sustainability explicitly, asking about any high-level sustainability initiatives or procedures they were aware of at their company, and utilized follow-up questions to uncover links (if any) between these high-level corporate goals and their role at the company. We concluded by soliciting a discussion of their personal perspectives regarding the EI of electronic devices and the factors and processes they desired to change in the future. 
\subsection{Participants}
We recruit two distinct categories of participants: \textit{LCA professionals} and \textit{engineers}. 
We define LCA professionals as individuals who work primarily to estimate the EI of products. These include analysts who directly estimate the environmental impact of devices as well as individuals who create analysis software and datasets to support product LCAs. 

We define engineers as individuals who contribute to the design, implementation, and manufacturing of computing devices. These individuals have responsibilities including designing physical hardware blocks, firmware implementation, and other design decisions that are manifested in end products. We also include individuals who oversee this process in a management capacity as part of this group. A breakdown of participants and a description of their relevant experience is shown in Table 1. 

Given that many companies in the consumer electronics space are highly protective of trade secrets and other intellectual property, we provided examples on how their responses would be anonymized in any resulting publications, such as removing any names of specific companies or products that could be used to infer their employer from included quotes. Participants were informed that we did not wish for them to share any information they felt may be sensitive, and were encouraged to describe representative scenarios rather than sharing potentially restricted details such as specific component numbers or internal product specifications. 
%added the description snowball sampling below. I'm not sure if there's a better way to describe our recruitment process in more detail? Adding the email template to supplementals might make them happy? 
We recruited participants through \review{snowball sampling, beginning by reaching out to }primary and secondary contacts within our professional network via e-mail\footnote{\review{Our email template for these initial emails can be found in the supplementary materials accompanying this work.}}. To aid in recruitment, participants were offered a \$40 USD electronic gift card for participating in the interview. Prior to recording, participants were provided with a verbal description of the study procedure and verbal consent was obtained. A written copy of the study procedure was made available upon request. Before recruiting participants, the study protocol was submitted to and approved by the institutional review board at the host institution.  

\begin{figure*}[t]
    \centering
    \includegraphics[width=\linewidth]{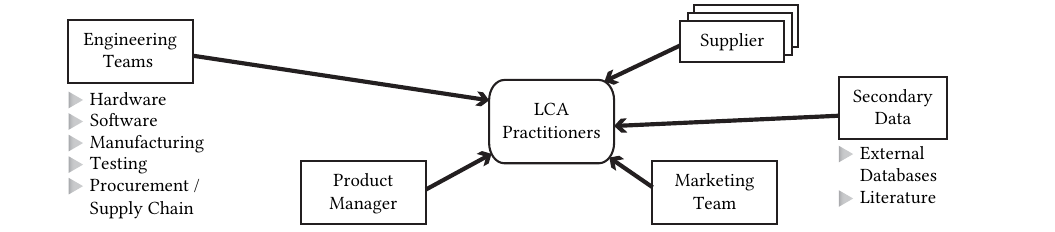}
    \caption{\textbf{LCA data collection.}
    LCA practitioners require input from many product stakeholders to collect the data needed to accurately model EI. 
    Without being part of the initial design process, they often must retrace the steps it took to arrive at the final product assembly.}
    \Description[LCA practitioners are in the middle of other stakeholders and receiving arrows.]{Engineering teams (including hardware, firmware, software, manufacturing, testing, equipment procurement, and supply chain management), Suppliers, Secondary data (including external databases and literature), marketing team and product managers are giving LCA information to LCA practitioners in the middle.}
    \label{fig:data_collection}
\end{figure*}

\subsection{Data Collection and Analysis}
All interviews were audio-recorded and transcribed using Zoom teleconferencing software for later analysis. These transcripts were first analyzed via open-coding\footnote{\review{Our codebook can be found in the supplementary materials accompanying this work.}} by two researchers separate from the interviewer utilizing the recording transcripts, with the raw audio recordings serving as a fallback to resolve occasional transcription errors. The study team then analyzed the codes using thematic analysis \cite{braun_using_2006} to understand current practices regarding LCA in the ICT industry, how the results of these assessments are (and are not) utilized when making product design decisions, and obstacles that make leveraging environmental impact information as part of the design process challenging. 

\review{We group the themes that emerged from our analysis into four categories centered around present LCA practices, the intersection of sustainability and product development, challenges specific to the ICT industry, and motivating incentive structures. We discuss our findings in the following section.}

\review{After concluding this process, we conducted member-checking with one randomly-selected LCA professional and one randomly-selected engineer, utilizing the four-question structured member checking interview procedure described in McKim~\cite{mckim_meaningful_2023}. Each participant was provided with a draft of the findings section of this paper (Section~\ref{sec:findings}) and invited to participate in a brief 10-minute interview to elicit feedback. In these interviews, both participants expressed that their thoughts and experiences were accurately captured in our analysis. }

%Alternatively, we could list the themes here:
% After conducting thematic analysis, we centered our findings around four central themes:
% \begin{samepage}
% \begin{enumerate}
% \item Present practices for LCA and how practitioners conduct these analysis on ICT devices and systems.
% \item Relationships between sustainability and engineering product development.
% \item Systemic factors impeding sustainability-oriented decision-making in the ICT space. 
% \item Incentive structures that motivate LCA and resulting impacts on stakeholders. 
% \end{enumerate}
% \end{samepage}

\section{Findings}
\label{sec:findings}
% We group the themes that emerged from our analysis into four categories.
\review{We begin by discussing} themes relating to the present practices for LCA and identify how practitioners conduct these analyses in the ICT space, identifying the process of gathering the necessary data from stakeholder groups as one of the most complex and time-intensive parts of conducting quality analysis. Next, we explore the relationship between sustainability and engineering product development, noting how engineers report struggling to relate abstract concepts of sustainability to their individual responsibilities. Third, we aim to identify the systemic factors that make sustainability-oriented decision-making particularly challenging in the ICT space. Finally, we examine the incentive structures that motivate LCA and identify how this impacts different stakeholder groups within companies.

%Removed above to move to 

\subsection{Current LCA Practices}
\label{sec:currently-done}
LCAs are essential for evaluating the holistic EI of products. For a basic definition of an LCA, see Section.~\ref{sec:lca-desc}. Below, we delve into the methodologies employed by LCA professionals and highlights the roles of stakeholder groups they interact with to conduct their analysis, including product engineering teams and external suppliers who provide the components utilized in end products. \review{We provide a visualization of the different internal and external stakeholder groups LCA professionals described requiring information from in Figure~\ref{fig:data_collection}. }

\subsubsection{LCA Within the Corporate Structure}
Both LCA professionals and engineers we interviewed described that companies rely on either a dedicated internal sustainability team (a similar structure to internal teams that verify security for products~\cite{poller-security}) or hire external consultants to perform sustainability analysis of products. As noted by LCA professionals and engineers:
\insertquote{The largest companies have in-house LCA practitioners, and most companies who've used LCA have hired consultants.}{\participant{9}}
%This sentiment was echoed by another participant who stated,
\insertquote{We have a team within the company that actually goes and tries to put a number to every component.}{\participant{11}}
%In addition to the perspectives of LCA professionals, this is further confirmed by the engineers we interviewed. An engineer described that
\insertquote{We have now an engineering team that is chartered with developing capability in that [LCA] area because it's kind of agnostic to what the product is.}{\participant{3}}

Regardless of the model, these quotes indicate that LCA is performed by individuals outside the product design team. This means that the LCA practitioner must actively seek and collect the necessary data to perform a thorough analysis.

\subsubsection{Collecting Internal Data}

\begin{figure*}[ht!]
    \centering
    \includegraphics[width=\linewidth]{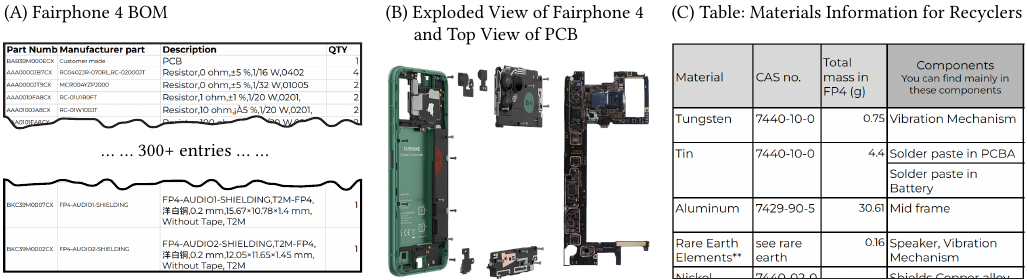}
    \caption{\textbf{Smartphone BOM.} (A) An excerpt from the manufacturing BOM for the Fairphone 4, which lists over 300 unique components. (B) An exploded view of the Fairphone 4 assembly~\cite{shcherban_fairphone_2023}. (C) Materials information for recyclers compiled and collected through the LCA data gathering process. Both A and C are from the manufacturer-provided repair and recycling information document for the Fairphone 4~\cite{fairphoneRepair}.}
    \Description[Two tables and an exploded view of a smartphone.]{The left side starts with an excerpt of the BOM of the Fairphone 4 listing each of the individual line entries,  followed by an exploded view of the smartphone assembly and a close-up of its PCB. Last, we have an excerpt of the materials information for recyclers, listing masses of different metals and the major components containing them.}
    \label{fig:fairphone_bom}
\end{figure*}

Performing an LCA requires accounting for the all of the material and energy inputs to the manufacturing process as well as the waste outputs. To do this, LCA professionals described obtaining a detailed BOM as an important first step. A BOM is a comprehensive list of raw materials, components, and assemblies required to manufacture a product. It includes detailed information such as part numbers, descriptions, quantities, and weights. \review{To illustrate this, we include an abbreviated example of the manufacturer-provided BOM for the Fairphone 4 smartphone, alongside an exploded view of device internals and the resulting materials information collected during the LCI phase in Figure~\ref{fig:fairphone_bom}. }

\insertquote{We need all the right data at this point. Like, we need your correct BOM with the correct weights and the correct descriptions of your part.}{\participant{7}}
However, this manufacturing BOM rarely contains sufficiently detailed information, such as the masses of raw materials and specific manufacturing methods, that LCA experts need to translate manufacturing specs into EI numbers. LCA professionals often go through an extensive process of researching information to fill in this missing information. 
%\insertquote{If you just ask people to give you a bill of materials, there will be probably things in it you don't understand}{\participant{1}}
%This process is further complicated by the need to understand individual specifications of components that are often not included in the part numbers and short descriptions present on a manufacturing BOM.
\insertquote{You never know what type of part this is, right? You don't know whether its injection-molded, whether it's bended, whether it's thermoformed, whatever, you always have to check another source of data which might be the technical datasheet and... they are in the system but they are not there on first glance, and they might not be there from your export of your BOM file.}{\participant{7}}

Further complexity is introduced for BOM entries that are purchased in batches, such as solder paste, flux, and other consumables used in manufacturing. 

\insertquote{If you have, like, a certain small printed wiring board, and it [the BOM] says 500 grams of solder paste, that's obviously crap. But it might be the number from batches you buy the solder paste in... Some of this is intentional, and some of it is just badly maintained data in your systems.}{\participant{7}}

Clarifying this information can make up the bulk of the time it takes to complete an LCA, and is largely manual process of reaching out to internal stakeholders for clarification:
\insertquote{Filling [the BOM], this is the main part, and this can take up to one month if it's really easy or up to a year if nobody works with you... I have regular meetings to fill the use case data or the sales numbers.}{\participant{7}}

Timelines for completing data collection are inherently tied to the responsiveness of other stakeholders who hold key information. Moreover, while LCA practitioners rely on data in internal CAD libraries, project management software, and contacts with the engineering team when possible, they often need to work with other internal teams who \textit{"work on the infrastructure of procuring equipment, installing it, maintaining it closely"} and with the supply chain to reconcile \textit{"all that data that is necessary to compute"} [\participant{5}]. Although all these teams may still be internal to the company initiating the LCA, the growing number of stakeholders significantly increases the complexity and duration of data collection. 
This highlights the necessity for robust internal communication channels and well-maintained data systems to facilitate efficient and accurate LCA completion.
%Moreover, while LCA practitioners rely on data in internal CAD libraries, project management software, and contacts with the engineering team when possible, they often need to work with other internal teams who handle procurement, maintenance, or other functions to reconcile missing information. 
%\insertquote{I would actually be focusing on... engineering teams related to that... Folks that work on the infrastructure of procuring equipment, installing it, maintaining it closely, working also with the customer supply chain folks to figure out, ok, where are things coming from? So there was also a huge part of the exercise of gathering or trying to gather all that data that is necessary to compute... on a process based basis}{\participant{5}}

%Although all these teams may still be internal to the company initiating the LCA, the growing number of stakeholders the LCA practitioner must seek out to request information significantly increases the complexity and duration of the data collection process. 
\begin{figure*}[ht!]
    \centering
    \includegraphics[width=\linewidth]{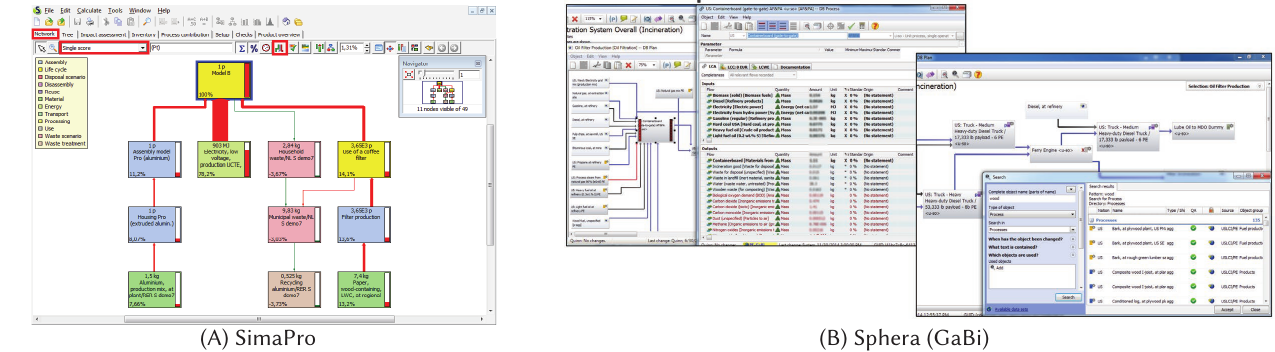}
    \vskip -0.1in
    \caption{User interface of industry-standard LCA software. (A) Screenshot of SimaPro LCA software from ~\cite{simapro-screenshot}. (B) Screenshot of Sphera (GaBi) LCA software from~\cite{gabi-screenshot}. We note that while these examples do not model electronics products, they are representative of each tool's user interface.
    }
    \Description[Screenshots of the two most widely used LCA software suites, SimaPro on the left and GaBi on the right.]{These representative screenshots show the interface of SimaPro and GaBi.}
    \label{fig:gabi-screenshot}
\end{figure*}
\subsubsection{Collecting External Data}
Many components in modern ICT products are themselves assembled from components produced by other companies, referred to in industry as "suppliers." However, these externally designed and manufactured components must still be characterized to conduct a complete LCA. 
Thus, LCA professionals must interact with suppliers to understand the EI of these external components. 

For \participant{9}, the best case scenario is if \textit{"you're getting a PDF by email from that supplier"} containing \textit{"a table showing the different impacts"}.
%\participant{9} starts by describing an ideal interaction with a supplier,
%\insertquote{Today, best case scenario, you're getting a PDF by email from that supplier saying, here's a table showing the different impacts we calculated for our product.}{\participant{9}}
However, suppliers often do not share the entire set of requested information that is needed to compute the LCA with acceptable uncertainty. 
A supplier might leave their own suppliers anonymized or \textit{"leave imprecision in specific quantities or specific material choices"} [\participant{9}]. 
%\participant{9} goes on to describe how
%\insertquote{The supplier might not want to tell their customers who their own suppliers are. 
%They might want to leave them anonymized, or they might want to leave imprecision in specific quantities or specific material choices}{\participant{9}}
Suppliers must currently respond to each request manually, creating an additional barriers for information sharing when request volumes are high for common parts: %especially when considering that suppliers for commonly-used components may have a high volume of these requests to process. 
\insertquote{So as a supplier you have to fill each [request] out. That's very time consuming and inefficient because we're basically putting the same information that they have into different types and shapes, etc.}{\participant{8}}
%LCA professionals have multiple strategies for dealing with situations when they are unable to obtain the required information from the product supply chain. 
When supplier data is unavailable, a common practice is to utilize pre-computed emissions factors for raw materials assemblies from databases such as ecoinvent~\cite{ecoinvent} and Sphera (GaBi)~\cite{gabi} or by referring to research publications. 
This type of data is referred to as secondary data. 
LCA practitioners describe the process of identifying which secondary data is the most suitable match for a given component or material as requiring substantial expertise and familiarity with relevant available datasets and current research literature. 
\insertquote{Which data do you use? Which secondary data will be the most suitable of the one? Even that that is quite complicated because you have this database... or you have some literature journal paper... so that opens up Pandora's box... People need to understand that it will not be trivial because you have to choose [which secondary data to use].}{\participant{2}}

This is particularly challenging for electronics where ICs may only be labeled by their physical footprint in databases such as ecoinvent~\cite{ecoinvent}. Unlike other industries where raw material costs often dominate, electronics manufacturing is both highly resource-intensive and variable. For example, all else being equal, a chip produced with a 7~nm process requiring advanced extreme ultraviolet (EUV) lithography will have a much greater environmental impact than an equivalently sized chip using an older 65~nm processes~\cite{gupta_act_2022}.

LCA practitioners occasionally resort to more drastic measures when stakeholders in the supply chain are unwilling or unable to share the desired information and secondary data is unavailable.
%When primary and secondary sources are unavailable, LCA practitioners occasionally resort to more drastic measures to gain information about the production of inputs to the product they are analyzing in situations where stakeholders up the supply chain are unwilling or unable to share the desired information.
\insertquote{Eventually the full-blown version is you go there and you make a true on-site verification... you can't get in and have a look at their accounts, but you can see how many trucks [are] running in and out of the place, and what kind of trucks, and what does it say on the side of those trucks?}{\participant{1}}
Ideally, collecting external data would be as straightforward as collecting internal data; however, communication with external suppliers often presents additional obstacles for LCA professionals.. %through selecting relevant alternative sources of data.

\subsubsection{Modeling EI}

After the data collection above, LCA professionals assemble an activity flow model, a graph-like structure mapping the material and energy inputs and outputs in LCA software such as SimaPro or Sphera (GaBi) as shown in Figure~\ref{fig:gabi-screenshot}. This is described by LCA professionals as much faster than data collection due to the structured nature of the flow modeling phase, which benefits heavily from the groundwork laid during data collection. 
\insertquote{If you have everything set up, it's rarely more than a few weeks of going back and forth having quality checks with your product manager or doing the quality check of what your associate did... so the data availability is the real pain.}{\participant{7}}

For small assemblies or processes the flow is often modeled manually using spreadsheet tools. However, for more complex assemblies or manufacturing processes, LCA professionals use dedicated LCA modeling software. 
\insertquote{We basically store our data in very simple files, often, it's just a CSV file and then we [use] fairly simple matrix inversion tools... if you have a fairly small system... Most of our customers use something called SimaPro, which is super old software... made back in the 1990s and they haven't really improved it since.}{\participant{1}}

These quotes highlight that once all the data has been gathered and reduced to quantities such as mass of a material or intermediate carbon footprint, producing an output becomes as simple as multiplying them to do unit conversions and summing up the output. Beyond database integrations and uncertainty calculations in some cases (see below), these tools do little to address the core challenges of data gathering described above. Moreover, their primitive interfaces are designed for a single expert user, making it difficult for engineers or others to collaboratively contribute to the process.

\subsubsection{What Makes a Good LCA}
%Responses about transparency, amount of information, what databases, uncertainty etc.
During our interviews, LCA experts shared with us the main quality indicators they use to judge the credibility of an LCA report.
One of the first indicators is the reporting of data sources.
\insertquote{First thing I would look at which database did they use. And knowing how few databases are actually mass-balanced, that gives my first kind of impression. Is this likely to be a serious study?}{\participant{1}}
Mass balance is the idea that the total mass of materials input to the process should be equal to the mass of the outputs. 
Ensuring that an LCA is mass-balanced shows that all parts of the life-cycle have been accounted for, even process waste and excess material.

Another important quality indicator for databases is their uncertainty modeling.
Ecoinvent represents uncertainty via log-distributions [\participant{2}], a common probability distribution in the LCA industry~\cite{heijungs2024lognormal}.
However, many databases do not contain this kind of information:\textit{``The big problem is currently that very few databases contain any reasonably high quality, uncertainty data''} [\participant{1}].
In practice, to improve uncertainty information, practitioners \textit{"give a worst case and a best case estimate"} [\participant{1}] and gather more information to lower that bound.

%\insertquote{you include the uncertainty on your data. And and that basically means if there's something you don't know about you. You give a worst case and a best case estimate, and then later, you use that to see, I mean, does it matter or not, and sometimes it doesn't matter, and then you go with it. If it does matter, you need to go and find better data.}{\participant{1}}
\participant{1} also expressed strong opinions on modeling by-products, which occur when an activity produces several distinct outputs.
This seems to be a limitation in current LCA modeling tools and hints at the use of different modeling strategies across LCA teams.

In general, interviewees said that transparency was one of their most important quality indicators.
A report should clearly show \textit{"potential limitations"} [\participant{7}] and \textit{"the flaws every LCA study has"} [\participant{7}].
It should be clear \textit{"who really did the study. Sometimes it's published by a company, and then somewhere in the imprint, you notice it was external"} [\participant{7}].
\participant{9} expressed regret over a lack of transparency and relying solely on \textit{"third-party brands to show the integrity of the data"}, which makes it difficult to \textit{"scale up the quantity"} and to reuse data for other studies.

%\insertquote{You know what the data is good for or not, as opposed to a culture where we only publish the final results, and you really have to trust us. But we went and got a bunch of a recognizable 3rd party brands to show the integrity of the data. That approach is too cumbersome to scale up the quantity and also the data available and get it integrated into more stages}{\participant{9}}
%\insertquote{you can look at how they treat byproducts, for example, that's also a major mistake that people tend to think that.}{\participant{1}}
%\insertquote{how transparent are there with potential limitations of your study.}{\participant{7}}
%\insertquote{they don't try to hide the flaws every Lca study has}{\participant{7}}
%\insertquote{description of data and who really did the study. Sometimes it's published by a company, and then somewhere in the imprint, you notice it was external}{\participant{7}}
Lastly, transparency is also valued for justifying why a new product had a significantly lower or higher reported carbon footprint that its competitors [\participant{7}], echoing the need for more comparability between LCAs.
%\insertquote{you want to use it and you might try to check first, carbon footprint per weight, for example, and then compare this to to similar products or similar data sets? whether they are playing in the same range. And if they're totally below, do they state this? ... they have green energy for a certain energy intensive technology or they are higher because they tried to do bit better quality. But they needed to have had higher scrap rate. But if you can really check why it's different, why it's super bad or super good}{\participant{7}}
%\textcolor{red}{not sure what to do with this section here, maybe remove the last simapro quote and add a quote about data quality including uncertainty being a sign of a good LCA, then conclude talking about how important this data gathering process is to conducting a LCA, and it's hindered by the LCA professional needing to initiate all of these conversations}

\takeaway Our interviews show the major challenge for LCA professionals is gathering the data for their analysis. This is often due to a lack of communication or understanding of the data needs by engineering teams and external suppliers. Tools and standards to support this are virtually non-existent, making it a time consuming manual task.

\subsection{The Intersection of Sustainability and Product Development}
\label{sec:engineers}
The section above reflects the challenges of sustainability experts performing the LCA. Next we explore the engineering and product design perspective, examining the challenges they face trying to incorporate sustainability into their designs and meeting these demands.
\subsubsection{Distributed Decision Making}
A finished product is an amalgamation of discrete choices on performance characteristics, feature set, components, materials, and manufacturing methods. Each individual decision has some EI associated with it. When asking engineers about their decision-making processes, we find it highly distributed across individuals and organizations. 

\insertquote{There's a team of leadership across various disciplines, so like hardware lead, software lead, manufacturing lead, test leads and stuff like that... Then they have a hierarchy below them of different teams focused on different features or different parts of the validation or engineering or design.}{\participant{16}}

%\insertquote{There's a team of leadership across various disciplines, so like hardware lead, software lead, manufacturing lead, test leads and stuff like that... Then they have a hierarchy below them of different teams focused on different features or different parts of the validation or engineering or design... there's a bunch of different disciplines that kind of lead up in a hierarchical fashion}{\participant{16}}
\participant{12} provided additional context by recounting the process of choosing a specific component:
\insertquote{Say I want the specific sub component, right? And then... the GSM [Global Supply Management] leader will want to use these specific suppliers, but maybe somebody else working on the software side doesn't like the reference code... or somebody else from hardware sees that, yes, like, of course the unit price is low but then you have all these extra passive components that make the system very complicated, or somebody may not like their EMC [Electromagnetic Compatibility] spec... There's a lot of people that are involved and nobody controls everything.}{\participant{12}}

The distribution of these design decisions makes it challenging for an individual to make sustainability focused choices. When optimizing a part for lower monetary cost, even project leaders must balance conflicting requirements across the organization suggesting optimizing for carbon cost would face the same challenges.

\subsubsection{Propagating Design Decisions to Final Products}
All engineering participants described following some form of phase gate design process utilizing design for manufacturing (DFM) principles~\cite{dfm-book, khurana2020beyond}. In this process, a list of specifications and features are developed and a series of prototypes are built and evaluated against these specifications. After prototype builds, multiple low-volume manufacturing test runs are conducted to validate the design and manufacturing process. However, we find that the bulk of architectural and component-level decisions are made in the earlier prototyping stage. When comparing the BOM of prototypes before beginning manufacturing validation testing to the resulting finished product, \participant{14} describes that 
\insertquote{Most [components] will make it in. After the proto builds, unless there's a subsystem like tear up or something due to some functional thing not being right, but usually it's not the case.}{\participant{14}}
This suggests that, from a temporal perspective, if sustainability-related metrics are to be considered and incorporated meaningfully into the design process, they must be introduced in the prototyping stage before choices have been solidified. 

\subsubsection{Translating Abstract Goals to Design Requirements}
While all engineers we interviewed were aware of high-level sustainability goals at their companies, they often struggled to understand how these goals translated to their role and responsibilities. 
\insertquote{When we have our yearly trainings they usually will have a snippet in there about... how we want to go towards sustainability. But a lot of times that's just like what the company in general is doing... It's not like we had a meeting to say hey, let's make this sustainable.}{\participant{10}}
\insertquote{I would say I don't have particular guidance on how this affects me daily. Maybe it affects... how the supply chain pre-qualifies suppliers?}{\participant{12}}
These quotes demonstrate that although engineers are aware of these high level goals, there is little awareness of how they are implemented or who is taking ownership to achieve them.
\participant{16} identified that goals must be expressed explicitly as measurable outcomes in order for them to be meaningfully considered within the design process:
\insertquote{The only way that you can really get some traction is like you have to be able to measure it and prove it, because if you kind of just generally gesture at it and don't have a measurable improvement or change or something, then you can't really include that in your calculation... as a KPI or key performance indicator, like product requirements.}{\participant{16}}

This is further supported by the fact that engineers who were more readily able to discuss sustainability within the context of their work consistently related it back to quantifiable metrics. \participant{4} described that in their role, this manifests itself most clearly through optimizing the power consumption of designs.
\insertquote{We want to design our chips to be low power, so that is tied into sustainability.}{\participant{4}}
\participant{3} went a step further, explicitly stating that they start by estimating the potential impact of various system inputs and outputs, then optimizing to reduce the factors that have the highest impact. 
\insertquote{The resources... water... power... gasses... the amount of energy required... so we try to monitor now all that and convert these to CO2 equivalents... We try to have the discipline to really understand what the drivers are... and if we can solve those items where the priority usage is... it becomes something that's real.}{\participant{3}}
Although this distinction may seem subtle, we note that this translates the abstract goal of reducing environmental impact into a requirement to reduce specific measurable product attributes, such as power consumption or gas usage. This allows engineers to factor sustainability metrics into the design discussions and workflows discussed above. 

\takeaway We observe that sustainability goals are often too abstract for individual engineers, and that decisions to prioritize sustainability must operate within many constraints. Goals need to be translated into familiar metrics that individuals and teams can use to take action.

Now that we have analyzed the perspectives of the LCA experts and engineers, next we zoom out to examine the ecosystem they work in, including the unique challenges of electronics and attempts to overcome them.
%\todo{Maybe identifying and overcoming challenges? Some of the things we kinda want to talk about here are just 'things that are hard' and not necessarily things that someone is working on improving}
%So far, we have mentioned themes which describe the way sustainability is currently handled, without it being a problem for the industry. 
%But where do people feel like they're encountering obstacles?
%This is a bleak picture for LCAs, but we mentioned that this is different for other industries. However, other industries have had success incorporating LCA into the design process. Why are computing devices especially hard?\todo{this above is a mess copy-pasted from removed section, need new intro}
\subsubsection{Navigating Supplier Relationships}
Computing devices are complex assemblies consisting of hundreds of parts from multiple different suppliers~\cite{lu_ecoeda_2023}. The varying power dynamics of supplier relationships can make procuring information uniquely challenging. 
\insertquote{That depends on your relationship to that specific supplier ... could be that they don't answer because they are super big, and they just don't care.}{\participant{7}}
This is further complicated by the multitiered supply chain for electronic devices. A component purchased from a supplier is frequently assembled from other components in turn manufactured by other suppliers. This is often discussed through supply chain tiers, where a tier 1 supplier is an entity directly purchased from, a tier 2 supplier supplies tier 1 suppliers, and so on.  
\insertquote{if you get past that first tier one supplier, anything beyond that usually is very, very difficult, because you have zero leverage also over that upstream tier.}{\participant{5}}
Since suppliers in tier 2 and beyond do not have a direct customer relationship with the company, they have little incentive to respond to requests for information. To further complicate matters, many companies in the ICT space consider other ICT companies among their customer base~\cite{tse2016embracing}, requiring companies to act simultaneously in both roles.  

\subsubsection{Intellectual Property Concerns}
Due to factors such as fierce competition for market share, reliance on similar groups of suppliers, large research and development (R\&D) costs, and the difficulty of enforcing patents internationally, companies in the ICT space place a high priority on safeguarding intellectual property (IP). Many suppliers fear sharing \textit{any} information about manufacturing and production practices for LCA fearing it could inadvertently leak critical trade secrets. 
\insertquote{IP is definitely the number one concern that is quite consistent throughout all the suppliers, because this is something that they are making a business off. So they do not want to go into the details.}{\participant{8}}
\insertquote{Another [challenge] is data security. So companies not wanting to expose trade secrets about their products externally.}{\participant{9}}
\participant{1} described that they often work with suppliers who have intellectual property concerns by requesting aggregate data.  
\insertquote{We say, just aggregate the things that you feel are confidential. So if there is specific chemicals, don't tell us about it. Just tell me how much chemicals are you buying and at what price... then I take a worst case assumption.}{\participant{1}}

\subsubsection{Technical Complexity}
Development of modern ICT products also requires integrating multiple distinct and technically complex components. Because of this, engineers generally specialize in a narrow application space. \participant{17} who works to implement verification logic on semiconductor devices describes how understanding the intricate tradeoffs of even highly-adjacent fields can pose a significant challenge:
\insertquote{There are a lot of other fields, say, like layout, like physical design... if they explain their problem they're facing right now it might take me a while to understand it, and not to mention that if I want to look at their work and then make a modification... it will take an infinite amount of time.}{\participant{17}}
Given that even specialized engineers experience this burden, it is surely magnified from the perspective of LCA professionals who often lack the field or industry-specific knowledge required to interpret the purpose and implied specifications of components when presented with a BOM.
\insertquote{It's super hard, because you have no idea what exactly this part should be...If you're a product engineer, you can check for plausibility of that specific product.}{\participant{7}}
\participant{7} goes on to describe the hypothetical example of encountering a part on a BOM called "left chamber." While this terminology may be familiar to an engineer working in the field and imply characteristics about its construction and function, a LCA professional conducting an assessment lacks the necessary context to interpret this information and must in turn request clarification from engineering teams working on the product. 
Just as LCA professionals express lacking necessary context to understand engineering documentation, we find that the inverse of this is also true. \participant{14} describes struggling to understand the impact of the environmental claims they see on product environmental impact reports. 
\insertquote{it's really hard to say. Oh, we use 30\% less water in production of this compared to what it was before. Like, I really don't know whether that's good or not. I mean it sounds good, but I don't know... what impact that has.}{\participant{14}}

\subsubsection{Scaling to Meet Growing Demand}
%Due in no small part to the factors identified in Section \ref{sec:motivating-factors}, 
\participant{5} describes there is a growing need for LCA: \textit{``There's a growing demand for people that can actually just carbon account at this point, because it hasn't really been automated"}. This is a positive sign but introduces challenges when scaling to meet demand due to the substantial domain-specific expertise required \participant{1} notes:
\insertquote{because the field is growing so quickly, so many new people are coming in that have absolutely no understanding of economics and stuff like that.}{\participant{1}}
This highlights an urgent need for better tooling to support LCA efforts, both to enable existing professionals to work more efficiently and to enable a wider group of participants to perform quality analysis. 

\subsubsection{Unclear Standards}
Several LCA professionals expressed mixed opinions about industry-agnostic LCA standards such as ISO 14040 \cite{iso14040} and ISO 14044 \cite{iso14044}. Although thorough in terms of defining what a LCA should include and consider, they do not specify the industry-specific implementation details such as software and data sharing practices. \participant{1} observes a need for a more concrete, application-focused standard for computing, saying:
 \insertquote{I think more and more people are beginning to realize that there's something wrong with the way most LCAs are done today and that we need to get a better standard, a better procedure, something and definitely more easy to understand.}{\participant{1}}
\participant{8} observes that standardization has already occurred in other industries
\insertquote{There are a lot of other industry that have already been through all this and developed their platform and their standardization of calculation method ... the chemical industry, the apparel industry, and the automotive industry...}{\participant{8}}
As an active participant in an industry group working to define standards for LCA methods in the semiconductor industry, \participant{8} describes that
\insertquote{the goal for that group is to actually standardize 2 things. One is the technology or platform that people use to exchange data and store data, and number 2 is to standardize the way that people calculate their carbon footprint, specifically, product level emissions.}{\participant{8}}

These quotes mirror the confusion felt by engineers and demonstrate a widespread recognition of the need for standardization. Designing new tools to facilitate the LCA process could both inform standards development and accelerate their adoption. 

\subsubsection{Possible Solutions}
Through the process of conducting these interviews, multiple participants raised specific ideas for ways the current state of the art could be improved upon. 
\participant{9} described a vision for digital collaborative LCA platforms:
\insertquote{I think it's moving LCA into a digital format where it can actually be used as a design tool and collaboratively between organizations is one of the key innovations that has to happen, because right now it's mainly a verification and accounting exercise outside of the individual people who are doing the detailed modeling work.}{\participant{9}}
By enabling more rapid computation of EI, \participant{9} hopes that building more flexible and collaborative tools could make the results of a LCA more accessible to stakeholders in the product design process. While tools like this could help address the temporal mismatch that exists between the rapid pace of product development in the ICT industry and the current LCA timeline, this alone does not address the challenge of providing actionable metrics to engineers described in Section \ref{sec:engineers}. Engineers expressed that, if integrated within their EDA and CAD software, they may be able to more easily consider sustainability-related tradeoffs. 
\insertquote{When you do your BOM, it shows you different things... You should have a column for sustainability, and like, if each component could have a footprint like you see when you make flight ticket bookings.}{\participant{15}}
\insertquote{Once it's reached that point of like oh, there's 10 components that match my spec, usually... I'll just go with the cheapest one, right? But then, if you can now see cost along with carbon footprint, maybe the decision changes. Maybe I'll spend, I don't know, 5 or 2 more cents on something and go with the part that's... better in terms of greenhouse gas emissions.}{\participant{13}}

\begin{figure*}[hbt!]
    \centering
    \includegraphics[width=\linewidth]{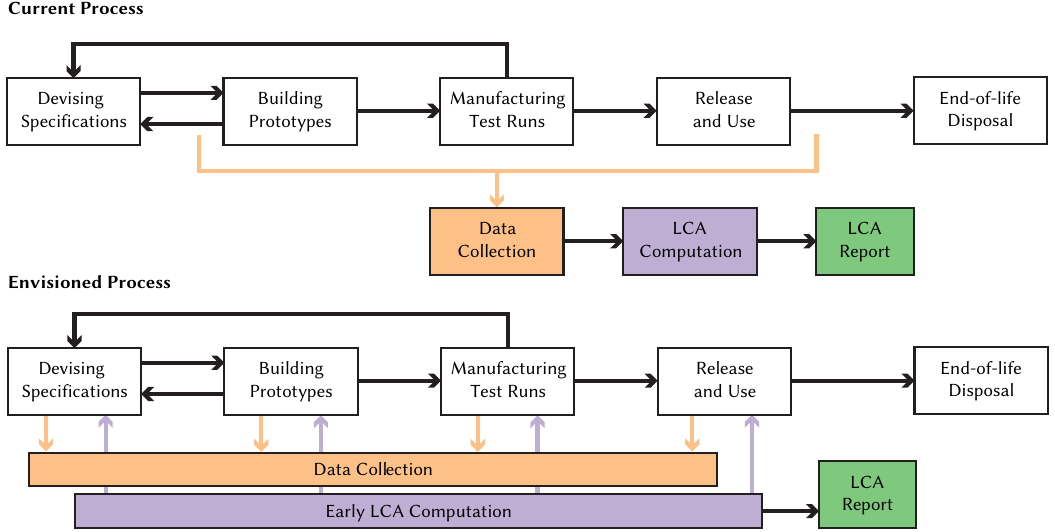}
    \caption{\textbf{LCA integration with product development.} 
    (Top) In the current product development process, LCA data is collection begins after the final BOM is completed. This means products have already reached the market by the time the LCA report is available. 
    (Bottom) In our envisioned product development process, LCA data collection and computation are carried out in parallel to all design stages in an integrated fashion. 
    The preliminary LCA results can now be used to inform decision-making.}
    \Description[Two rows showing product development and LCA report timeline alignments.]{
    The top row shows five product development stages in black and white. 
    In color, it shows the LCA report stages, which are shifted to the right with respect to the product stages.
    The bottom row also starts with five product development stages in black and white. 
    In color, it shows the LCA report stages, which are aligned with the product stages.
    Arrows from the LCA computation box go back to the product development boxes.}
    \label{fig:timeline}
\end{figure*}

\takeaway In addition to the challenges faced by LCA experts and engineers, the current ecosystem around them is not conducive to data sharing and there is a lack of accepted standards or tools. There is however clear momentum and interest in adopting such solutions \review{to more tightly integrate the LCA and product design timelines. We visualize the current process and a future, integrated timeline in Figure~\ref{fig:timeline}.}

\subsection{Incentive Structures Between Stakeholders}
%In the previous sections we seek to elucidate the current process LCA practitioners use to conduct their analyses, how product engineers engage with this information, and systemic factors that make incorporating LCA information into the design process challenging. 
In addition to exploring how different stakeholders engage with LCA and the design ecosystem around them, it is also important to understand the incentive structures present within companies and the broader ICT industry.
%and how LCA fits within the broader context of companies working to design products and services within the ICT industry. 
\subsubsection{Motivating Factors}
\label{sec:motivating-factors}
Participants report government regulations, negotiations to secure environmental resources, and corporate net-zero goals as the main incentives motivating LCAs. All LCA practitioners explicitly mentioned current and anticipated government regulations and reporting requirements as a central driver for conducting life-cycle assessment. 
\insertquote{[motivation] definitely comes from the legal or regulation part of the world, where, as you may already know, there are different kinds of reporting regulations that is coming up both in the US but more aggressively in Europe.}{\participant{8}}
As an example of this, multiple participants mentioned recent EU regulatory framework requiring LCA and EI reporting for products containing rechargeable batteries~\cite{battery-regulation} and the EU Digital Product Passport (DPP) initiative\cite{DPP-EU}.  
Incentives also come from negotiations with local or regional governments who oversee natural resources and land management. Modern semiconductor fabrication facilities have increasingly expanded in scale to enable increased efficiency \cite{yeung_explaining_2022}. The scale of these operations means that manufacturers must secure non-trivial agreements for land use, energy, water, and waste management from local governments. \participant{3} describes conducting LCAs as an important component of negotiating for approval to expand manufacturing capacity in a region. 
    \insertquote{If they want to expand their facility and the town they're in says `we don't believe you're a responsible partner, we're not going to let you use more power... more water' Now they have to go find another place to build their next factory. So it's becoming an economic necessity because the alternatives are very, very costly.}{\participant{3}}

On an industry level, \participant{8} pointed out that the semiconductor industry needs the ability to compare with other industries both to conform with reporting standards as well as to further garner support from suppliers and manufacturers for LCA processes. 
\insertquote{A baseline is important, because in order to speak at an industry level and gather traction in in terms of resources and support to reduce carbon emissions, we have to first understand what is the current emissions of the industry, right? Without that we cannot compare with other industries.
}{\participant{8}}
The final motivating factor, mentioned by both LCA professionals and engineers, were the numerous carbon reduction goals pledged by companies in the ICT space. 
    \insertquote{They have very ambitious net 0 goals. In order to achieve that, they need more granular data to support them on quantifying initiatives, decarbonization efforts, and quantifying the roadmap so they can make it more credible.}{\participant{8}}
    \insertquote{They're quite public. I think by 2030 we want to be net 0 total emissions... [this is] what we promise to everyone including our shareholders but also our customers.}{\participant{11}}
We note that all of these stated goals are in terms of carbon emissions, which is in contrast with the perspectives shared by LCA professionals who spoke in terms of EI more holistically. We could not infer any tension between stakeholder groups arising from this difference perspectives in our interviews, but this is nonetheless illustrative of the high stated priority of carbon emissions reduction relative to other EI metrics. 
\subsubsection{Value for Stakeholders}
These economic and regulatory forces create incentives and value for sustainability in the industry. However, it is not immediately clear how these high-level goals impact stakeholders at a more granular level. We next examine the stakeholders who are impacted by an increased emphasis on sustainability as a metric. 
Engineers and LCA professionals both cited marketing teams as a dominant internal advocate motivating discussion of sustainability within the context of an individual product and for collecting LCA data. %Both engineers and LCA professionals identified marketing teams as a driving force for collecting LCA data on products. 
\insertquote{A marketing person might go to a VP and say, hey, we want to have a marketing sticker that say's it's sustainable... in which case they might tell the MechEs [mechanical engineers] hey, you have to design with this specific material.}{\participant{10}}
%\participant{7} described this more succinctly, stating
\insertquote{They want to launch and then use that [LCA result] for marketing.}{\participant{7}}
While marketing teams may derive a benefit from being able to report product sustainability metrics in materials, current processes exist largely as a burden for other stakeholders, including suppliers and the teams building these products. 
Individual suppliers struggle with these added requirements to report sustainability information, especially since they may not have dedicated teams to respond to requests to information from LCA professionals. 
\participant{8}, an industry consortium manager focused on developing LCA standards within the semiconductor industry, went on to clarify that the value proposition of engaging in this process with customers is often unclear from the perspective of suppliers. 
\insertquote{In the end of it that does not generate a lot of value for suppliers. So it's basically just a requirement for additional work.}{\participant{8}}
This sentiment was also echoed by \participant{5}, who noted that
\insertquote{5 to 10\% of suppliers either have the data readily available or are keen to provide that quickly to you. The rest of them either has no capacity, or doesn't have funding available to procure that data, or doesn't even know where to get it often or sometimes also just is not really interested in providing it to you because there's no financial benefit for them yet.}{\participant{5}}
Product engineers also describe feeling similarly burdened, explaining that managing sustainability-related requirements further complicates their already challenging design processes. 
\insertquote{There's already so much work involved in the design of a product, and having to think about ways to navigate around recycled goods or recycled materials... It is kind of an afterthought... when you're developing prototypes and trying to get something to work.}{\participant{13}}
\participant{3} further emphasized that other product requirements are not relaxed when adding sustainability-related design requirements:
\insertquote{There is no trade-off... we don't get to work on sustainability in lieu of some other need or requirement so it becomes yet another requirement.}{\participant{3}}
This outlines the different stakeholders involved and their value perception of sustainability. While corporations stand to benefit in terms \review{marketing and compliance with regulations, others such as suppliers and individual engineers are not currently equipped to handle this additional workload. }

\takeaway Government regulations, economic forces, and marketing are key incentives for sustainable design; however design teams are often not involved in making decisions and lack tools to support sustainable design, making it an additional burden.

\section{Opportunities}
\label{sec:opportunities}
\begin{figure*}[ht]
    \centering
    \includegraphics[width=\linewidth]{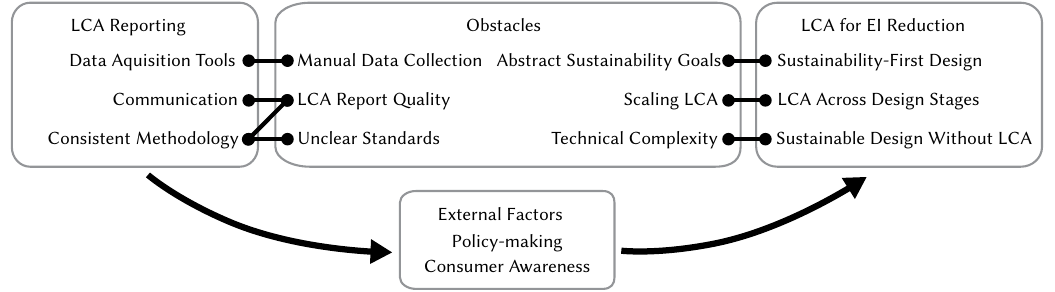}
    \caption{\textbf{Mapping obstacles to opportunities.} \review{Here, we tie opportunities for innovation to obstacles identified through our interviews with LCA professionals and engineers. Better LCA reporting could provide support for external factors such as policies and consumer awareness of sustainability, which may in turn motivate efforts in EI reduction using LCAs.}}
    \Description[The mapping between identified obstacles in Section~\ref{sec:findings} and opportunities in Section~\ref{sec:opportunities}.]{Obstacles include: (1) manual data collection (mapped to Data acquisition tools under LCA reporting opportunities); (2) LCA report quality (mapped to both Communication and Consistency in Methodology under LCA reporting opportunities); (3) Unclear standards (mapped to Consistency in Methodology under LCA reporting opportunities); (4) Abstract sustainability goals (mapped to Sustainability-First Design under LCA for EI reduction opportunities); (5) Scaling LCA to meet demand (mapped to LCA across design stages under LCA for EI reduction opportunities); (6) Technical complexity under (mapped to Sustainable design without LCA under LCA for EI reduction opportunities). External factors include policy making and consumer awareness.}
    \label{fig:mapping}
\end{figure*}
Based on the major themes that emerged from the interviews, we identify the following research opportunities on computational systems and interfaces to support the LCA process and to better incorporate sustainability in product development. We divide these opportunities into two main categories. The first is ways to improve the LCA process itself through automating data collection, improving data sharing, and methodology standardization. Second, we focus on opportunities to \review{leverage LCA information during design to reduce EI. A visual mapping of opportunities to the obstacles they address is shown in Figure~\ref{fig:mapping}. In the following sections, we elaborate on specific opportunities within each category and discuss how HCI research could help address current limitations while supporting more sustainable product development practices.}
% % breaking down LCA tools into LCA tools that make doing LCAs faster and easier, and tools that make LCAs more "right"
We also acknowledge that \review{efforts to reduce the EI of electronics are moderated at a high level by external factors such as regulatory policy and consumer choices. We hope that research in the HCI community can provide tangible support during the design phases, in addition to current work that seeks to} inform policy-makers and consumers to further incentivize sustainable product design.

%While we hope that this area of research can be useful, we acknowledge that technology and the HCI community alone cannot resolve systemic problems which are political in nature. 
%\todo{this may be unnecessary since it will be clear from the intro that the focus is on how CHI can contribute. let's see and iterate.} \todo{VI: I think we can put this in the discussion/conclusion}

\subsection{Facilitating LCA and EI Reporting}
\subsubsection{Data Acquisition}

The majority of the time spent producing an LCA is dedicated to data collection. This is due to the large volume of required data and the involvement of multiple stakeholders who use different data formats, computation and collection methods. 
We identify four opportunities to support faster data acquisition cycles:

\paragraph{Improved interfaces for data translation between stakeholders}
%We often observe that LCA information exists somewhere, but in a different software or under a different terminologies, used by other teams, leading to delays.
To facilitate data requesting and sharing between suppliers and companies, we can develop interfaces that can use the languages that stakeholders are most familiar with to communicate the needs, and a common underlying representation for the requested and shared data.
%Encouraging data sharing can potentially collectively reduce the work for companies in the ICT industry to do LCAs for their products.

\paragraph{Automating data collection}
Data collection can be automated with IoT devices, as suggested by \participant{2}, using readily available sensors for power usage and heat dissipation or computer vision techniques to track waste and excess material. While tracking devices themselves may contribute to EI, low-tech, passive sensing techniques offer a promising approach to mitigate this issue~\cite{zamora2024moirewidgets, johnson2023millimobile}.

\paragraph{Automating data approximations.}
Since LCA practitioners commonly use industry averages and approximations to compensate for missing primary data, a promising approach is to automate this process using data crawling techniques or AI-augmented search engines to discover available information~\cite{sluicebox}, combined with data-driven approximation algorithms.

% As it is common for 
% LCA practitioners  use industry averages and approximations to overcome missing primary data.
% However, finding these approximations still takes time.
% One promising avenue is to leverage data crawling techniques or even LLM-driven search engines to facilitate the discovery of available information~\cite{sluicebox}.
% Additionally, searching for product-related information, such as technical specifications or internal photos could be useful information to support establishing the BOM.
% \todo{I thought this was a bit too specific so changed to make it more general}

\paragraph{Targeted data collection for uncertainty reduction.}
While gathering more primary information is always beneficial for improving LCAs, it would be valuable to support LCA experts in prioritizing which data sources to target to reduce overall uncertainty. This can be non-trivial because the impact depends not only on how much uncertainty a given data source has but also on its position in the data flow. A promising direction is to frame this as an inverse optimization problem, modeling both the impact of each data source on the final uncertainty and the difficulty of data acquisition, with internal sources generally being less costly than external ones.

\subsubsection{Consistency in methodology}

A key challenge in conducting LCAs is to guarantee consistency across different assessments. 
LCA practitioners gather EI data from a variety of sources, including primary data from suppliers, secondary data from industry-averaged databases, and regression models from academic papers. 
These data sources are often aggregated in spreadsheets, with varying degrees of quality and uncertainty. 
Managing, maintaining, and reusing this diverse data is a significant hurdle. 
Additionally, LCAs are typically produced by separate teams, which can lead to inconsistencies. 
For example, \participant{2} described an experiment where two LCA teams independently produced assessments for the same product, and the outcomes differed by 30\%, highlighting discrepancies in modeling choices.  
%\todo{shouldn't we move this to 4 and keep this shorter?}
To address these challenges, we identify two key opportunities for improving consistency:

\paragraph{Collaborative tools}
We think that fostering exchange and enabling collaborations between multiple LCA practitioners with real-time collaborative tools could open the door for more methodological agreement.
In recent years, browser-based collaborative tools, such as Google Docs and Overleaf for writing, Figma and Canva for UI and graphic design or Onshape for CAD, have shown success for this tool paradigm. 

\paragraph{Heterogeneous database management}
Instead of enforcing strict EI data format standards, we propose developing systems that embrace the inherent messiness of real-world data sources. 
This approach, which lies at the intersection of computer systems and creative processes, could help LCA practitioners manage and integrate diverse data formats without sacrificing accuracy. 

\subsubsection{Communication}

An LCA is more than just a final number; it involves detailing product components, associated activities, and connecting them via activity flows. A major challenge, as noted by \participant{1}, is communicating the LCA model to stakeholders and non-experts. We identify three opportunities for future research to improve LCA communication:
%We need to clarify the methodologies and the metrics used in this process before it can be meaningfully interpreted by a reader. 
%it does not mean anything unless you give it the details of what you're measuring and how and this makes it hard because it's hard to interpret and it's hard to compare two things because you have all this explanation and these explanations can be different.

\paragraph{Explainability and contextualization}
%As mentioned by \participant{1}, it can be challenging to communicate LCAs with non-experts.
%To enable exchange with design teams, we need to find a way to share LCA results which goes beyond the final EI numbers.
%The fundamental challenge in communicating LCAs is their complexity.
LCA graphs can be complex, representing products at a low level, but they also reveal human activities that offer opportunities for visualization and storytelling.
We see potential in using hierarchical storytelling to emphasize the human aspect of LCA models, which otherwise tend to commodify human activities. As an example, program visualization methods could assist engineers in connecting their design decisions to the overall EI of a product.
%We see potential in using hierarchical storytelling and program visualization methods to enhance the emotional impact of LCA models, which otherwise tend to commodify human activities, as well as assist engineers in connecting their design decisions to the overall environmental impact of a product.

%\todo{I'm not sure what emotional impact  here means}

\paragraph{Comparisons}
%A major challenge in LCA is the comparison between different LCAs (\participant{2}, \participant{9}).
% Communicating one LCA is already challenging, but often we need to make comparisons between similar products or design variations, i.e., between different LCAs (\participant{2}, \participant{9}).
% %Comparing the final EI result is only possible under certain conditions such as a sufficiently low uncertainty and the same modeling assumptions.
% Understanding the difference between two LCAs graphs involves comparing their modeling assumptions, data sources, etc.
% Analogously to program comparisons, we want to perform an \textit{LCA diff}.
% However, program comparisons are challenging and while locally optimal solutions for moderate sized programs exist~\cite{rinaldi2023nodegit}, LCA specific comparison approaches, from semi-guided up to fully automatic approaches, would strongly benefit the community.
% Similar to program comparisons, we propose an \textit{LCA diff}. While program comparisons are challenging, developing semi-guided to fully automatic LCA comparison methods would greatly benefit the community.

Given the need to compare similar products or design variations (\participant{2}, \participant{9}), there is potential for LCA tools tailored to comparing two LCA graphs, which involves analyzing modeling assumptions, data sources, and more. This could draw from prior work on program comparisons, which are challenging. While locally optimal solutions for moderate-sized programs exist~\cite{rinaldi2023nodegit}, developing LCA-specific comparison approaches, from semi-guided to fully automatic, would be valuable.

\paragraph{Rethinking EI metrics}
During our interviews, engineering participants expressed that metrics such as the global warming potential (GWP) were not very meaningful to them.
While other metrics have been proposed by \participant{1}, such as quality-adjusted life years, conveying the effects of climate change and sustainability-related information is an ongoing challenge~\cite{jansen2023interdisciplinary} and leaves room for new approaches.
One possible alternative for displaying an absolute result is to display a relative result or ranking or to communicate context via related products or decisions, better aligning with the optimization-based decision-making approaches described by engineers in our interviews.  

\subsection{LCA to Inform EI Reduction}\label{sec:lcareduction}

\subsubsection{Sustainability-First Design}
%\subsubsection{Sustainability as an Additional Requirement}
Sustainability adds to design requirements like functionality, cost, and aesthetics, making integration more challenging. We propose three ways to incorporate sustainability into the design process:

\paragraph{Interpretable metrics}
One possible avenue is providing engineers with clearer, more actionable values to guide their decisions. Several participants noted that sustainability metrics are hard to interpret and desired simpler, more intuitive metrics, such as sorting components by carbon footprint on vendor databases or choosing parts with an eco-label (\participant{14}). Expanding to other impact factors like component toxicity would be valuable to optimize for e-waste reduction, but require addressing challenges in both estimation~\cite{jainelectronic} and communication, as toxicity includes various types of ecological and human impacts.

\paragraph{Trade-off visualization}
Viewing the design process as an exploration problem~\cite{hsueh2024counts}, we propose framing sustainable trade-offs as a multi-objective optimization problem. Decision-makers need visualization tools to explore the Pareto front of design alternatives. These tools should display various metrics and offer options while ensuring the design remains valid, which can be challenging~\cite{lu_ecoeda_2023}.
\paragraph{Constraint-based systems}
Rather than optimizing for sustainability, an alternative is to treat it as a constraint, imposing rules on the design space and guiding design choices. An example is the zero-waste garment design paradigm, where a constraint-based system ensures no fabric waste is produced during the process~\cite{zhang2024wastebanned}.

\subsubsection{LCA Across Design Stages}
% % you want to take into account LCA at every stage, in particular hgih-levle design decision really matter downstream and you want to be able to make informed decisions 
% % From the outset, our motivation was to consider LCA not just as an EI reporting tool, but as an EI reduction tool.
% % But what does it mean to design with LCAs?
% Designing involves different high-level and low-level exploration phases.
% We sketch, quickly develop 3D models and create physical mock-ups.
% What makes or breaks a design workflow is the ability to make changes, sometimes substantial, high-level changes.
% However, LCAs are a very low-level representation of a design.
% If we want to integrate, get informed by, and act upon LCAs during design, then we need to be able to edit effortlessly both the LCA and its product design level counterpart~\cite{}.
% Our three research opportunities to edit representations are:
Designing involves both high-level and low-level exploration, from sketches to 3D models and physical mock-ups. 
If we want to integrate, get informed by, and act upon LCAs during design, we need to be able to reflect high-level changes on low-level LCA representations.
To enable sustainable design choices earlier in the process, before the product is finished, we propose two research opportunities:

%Two solutions:
%- hierarchical approach - we model 

\paragraph{Representations at different levels of detail}
We think that one promising avenue for designing with LCAs is to maintain representations at different levels, such as low-level LCA and high-level product designs and to translate changes between them.
This translation problem can be viewed as a compilation problem, for example if all representations are different programs, or as a bidirectional editing problem, for example if one of the representations is a data structure~\cite{cascaval_differentiable_2022}. 
The challenge here is to find the right representation and translation methods for design.

% \paragraph{Inferring LCA representation}
% Instead of maintaining and adjusting the LCA model from a previous design phase, we could also automatically generate a new LCA model for each edit.
% BOM and graph generation methods are challenging problems, but future availability of more LCA data and searching for alternative data sources could prove a fruitful avenue.
%for example, if Imake this out of plastic vs wood, what which one will be better? tools that can infer from data 
%how these high level design decision will impact downstream. 

\paragraph{Hierarchical LCA modeling}
Starting an LCA can be intimidating because of its ambition to holistically capture all human activities tied to a complex product. 
As in other disciplines, LCA practitioners talk about bottom-up versus top-down modeling strategies to overcome the blank canvas.
However, current LCA design tools only work with low-level database entries, so-called processes, which are connected to flows to create complex LCA graphs.
We think that hierarchical LCA design tools that facilitate top-down approaches could benefit LCA practitioners.
Additionally, hierarchical models provide high-level editing handles by design. 
%\todo{I don't full understand this so couldn't summarize it}

\subsection{Sustainable Design Without a Full LCA}
LCAs are a powerful modeling tool that requires expertise and time investment.
But even without incorporating the full LCA paradigm into design tools, EI factors can be taken into account.
Here, we outline three opportunities to enable sustainable design as the ecosystem of LCA tools described above develops:

\paragraph{Streamlined LCA methods}
Streamlined LCA methods~\cite{olivetti_product_2012,sluicebox} do not establish a full LCA graph, but they directly produce a carbon footprint given product characteristics or a detailed BOM.
They are often used to get an approximation of the EI of a product and they might be useful to scale up carbon assessments in the light of upcoming EI reporting legislation [\participant{2}].
%Producing an LCA takes time for data collection and modeling.
%However, legislation such as the Digital Product Passport will soon require all products to report their environmental impact.
%According to our participants, the LCA industry is not ready for that amount of assessments.
%Streamlined LCA methods, such as PAIA~\cite{olivetti_product_2012}, Sluicebox, etc., could be a solution to this problem.
The challenge with these methods is their opacity and possibility for data quality inspection and uncertainty modeling.
%We think that while streamlined LCA methods might be necessary to scale the number of produced LCAs, interfacing with these methods and inspecting their results and computation functions could be a potentially interesting avenue.
%In parallel, we should work on making traditional LCA modeling workflows more accessible for a wider audience of practitioners.
We think improving them and developing interfaces to inspect their results and computation functions could be an interesting avenue for the community to explore.

\paragraph{Proxy-metrics}
Carbon accounting is relatively new compared to tracking other metrics, such as monetary cost and power usage.
Without reinventing an accounting model for EI factors, can we leverage existing factors as a proxy metric?
For example, is the manufacturing cost negatively correlated with carbon footprint?
Intuitively, cheap products are often associated with short life spans and high waste, but a cheap product could also hide a more sustainable, low-tech solution.
What insights can we gain if we add power consumption, heat dissipation, and water usage to this picture?
We think that a wide-ranging study of different, already existing metrics with respect to modeled EI factors could reveal interesting multi-variable correlations.
These correlation models could then be used for faster EI approximation methods during design.

%\paragraph{Working with parts libraries.}
%Supply engineers and PCB librarians work with internal component libraries, where each part has been vetted and even already been used in another product.
%Already existing expertise, such as developed firmware, is important and requesting the approval of a new part can take several months (P3).
%For EDA research, it is important to consider these practical limitations when developing tools.
%Searching for a functionally equivalent, but less carbon expensive part on online libraries, e.g. DigiKey, should be practically feasible, but its selection should be more expensive than the selection of a part of an internal library.
%\todo{another idea?}

\paragraph{Cross-pollination between different research areas}
An LCA translates a product BOM into human activities, which are converted into environmental impact.
However, human activities could also be converted to other factors, such as physical and mental labor, focusing on the people affected by a proposed life cycle.
This is a similar approach to research which proposes domain-specific languages for maker workflows~\cite{tran2024tandem} and laboratory experiments~\cite{li2022root}, guiding human activities in these spaces.
We see cross-pollination opportunities between LCAs and human activities modeling research which could present a gateway for makers and amateurs to explore these modeling paradigms for both environmental and human factors.
%We imagine the LCA paradigm being used for electronics makers who want to formalize their own workflow and analyze environmental impact factors and other custom aspects.

%We think that there are cross-pollination opportunities between LCAs and research which proposes frameworks and domain-specific languages around maker workflows~\cite{} and lab experiment setups~\cite{}.
%We want to think at a higher level about designing flows of human activities.\
%\todo{Maybe mention resources like SnapEDA and Ultra Librarian which provide crowd-sourced footprints for parts as a potential model?}

\section{Limitations}
While this paper provides valuable insights towards incorporating LCA more holistically within ICT product design, several limitations with our approach should be acknowledged. 
Our participants were overwhelmingly male (16/17) and all based in either North America or Europe, which limits our perspective within the context of the ICT industry, which is truly global in scale. This may limit the ability of our findings to generalize to other regions with different regulatory environments, market conditions, and attitudes toward sustainability. 

In this study, we focused specifically on two stakeholder groups, LCA professionals and product engineers in the ICT space, in order to identify opportunities to bridge the gap between assessing EI and designing to optimize for EI. However, as we illustrate in our findings, there are countless additional stakeholder groups that incentivize or participate in the product development process in the ICT industry. When discussing the role of these additional stakeholder groups in this paper, we do so through the perspectives of our interview participants. Future work could engage directly with these additional stakeholder groups to provide a more comprehensive and nuanced understanding of their involvement in the product development process. 

In addition, the confounding factor of social desirability bias is well-documented when engaging with topics such as environmental sustainability\cite{social-bias-roxas, factorial-survey-cerri}. Though it may be challenging in practice, this highlights both the need for and potential value of conducting observational studies on product development and sustainability in corporate contexts. 
By acknowledging these limitations, we aim to provide a transparent account of our study's scope and encourage further research to address these gaps and build on our findings. 
\section{Conclusion}

% A summary of what where the key challenges we identified that are relatable to CHI, I thought they would be nicely broken down into (1) People-people things, (2) multiple people managing data things, and (3) data being use to inform people decions making -> the reason I thought these 3 things were worth highlighting was because they are all very much CHI. I can't think of a 4th one but if there is you can add it too.
In this paper, we explored the obstacles and opportunities for improving the \review{ integration betwen the LCA process and }sustainable electronics design. Through a series of semi-structured interviews with industry experts, we identified several key challenges related to LCA acquisition and use. These challenges stem from three main areas: (1) how different stakeholders—LCA experts, product engineers, managers, marketing teams, and others—communicate with each other, (2) how data is shared and passed between these stakeholders, and (3) how this data is used to inform decision-making processes. These areas are rich with opportunities for HCI researchers, as improving interactions between people, between people and data, and designing tools that help interpret data are core aspects of HCI.

% this is basically a summary of what we identified in 5,
Building on these challenges, we identified specific opportunities where computational systems and interfaces can make LCA reporting more efficient, accurate, and interpretable. We also highlighted how computational tools can help integrate LCAs into the design process—by supporting design decisions in the context of competing objectives, enabling decision-making earlier in the design stages, and allowing for informed decisions even without perfect LCA data.

% this is a note that in addition to the opportunities that are directly actionable by Chi (section 5) there are the policy/consumer awareness things but that by building new tools, CHI people  can also end up helping to influence them as well. 
Additionally, our study gives insights into the complex incentive structures surrounding the adoption of sustainability practices, shaped by factors such as consumer awareness, policymaking, and the costs imposed on suppliers. While addressing these systemic issues requires broader changes, we argue that improved computational systems and interfaces have the potential to accelerate sustainability efforts. By making LCA processes more accessible, interpretable, and seamlessly integrated into design workflows, these tools can influence consumer, policy, and industry practices.

%endign with a call to action.
The CHI community, therefore, not only has the power but also the responsibility to drive these efforts forward and contribute to the broader goal of sustainable electronics production. We hope our work will inspire future research and the development of innovative systems that promote environmental responsibility within the electronics industry and beyond.

% In this paper, we explored the obstacles and opportuniteis to improving the integration between the Life Cycle Assessment (LCA) process and sustainable product 

\begin{acks}
The authors would like to thank Rushil Khurana and Richard Li for their guidance on study methods, as well as the anonymous reviewers for their feedback and suggestions to improve the manuscript. 

This research is based upon work supported by an Amazon Research Award and the National Science Foundation under award numbers CNS-2401177, CNS-2338736, CNS-2310515, and IIS-2212049. Any opinions, findings, and conclusions or recommendations expressed in this material are those of the authors and do not necessarily reflect the views of the National Science Foundation.

Vikram Iyer holds concurrent appointments as an Assistant Professor in the Paul G. Allen School of Computer Science and Engineering at the University of Washington and as an Amazon Visiting Academic. This paper describes work performed at the University of Washington and is not associated with Amazon.
\end{acks}
%TC:ignore
\bibliographystyle{ACM-Reference-Format}
\bibliography{references, more_refs}

%TC:endignore
% \appendix

\end{document}